\documentclass[sigconf,screen,9pt, 
dvipsnames,svgnames,x11names,table]{acmart}
\renewcommand\footnotetextcopyrightpermission[1]{}
\usepackage{glossaries}
\usepackage{xcolor}
\usepackage{xspace}
\usepackage{blindtext}
\usepackage{graphicx}
\usepackage{textcomp}
\usepackage{color, colortbl}
\usepackage[inkscapepath=./figures]{svg}
\usepackage{breakurl}

\usepackage{relsize}
\usepackage{cleveref}
\usepackage{hyperref}
\usepackage{subfloat}
\usepackage{multirow}
\usepackage{subfig}
\usepackage{setspace}
\usepackage[margin=5pt,font={stretch=0.9}]{caption}
\usepackage{url}
\usepackage{tikz}
\usepackage{fancyhdr}
\usepackage{tcolorbox}

\def\BibTeX{{\rm B\kern-.05em{\sc i\kern-.025em b}\kern-.08em
    T\kern-.1667em\lower.7ex\hbox{E}\kern-.125emX}}

\definecolor{mymauve}{rgb}{0.58,0,0.82}

\hypersetup{
  colorlinks=True,
  linkcolor={green!50!black},
  citecolor={red!70!black},
  urlcolor={blue!70!black}
}

\newboolean{showcomments}
\setboolean{showcomments}{false}
\ifthenelse{\boolean{showcomments}}
{ \newcommand{\mynote}[3]{
    \fbox{\bfseries\sffamily\scriptsize#1}
    {\small\textsf{\emph{\color{#3}{#2}}}}}}
{ \newcommand{\mynote}[3]{}}

\newcommand{\rk}[1]{\mynote{Rob}{#1}{mymauve}}
\newcommand{\fg}[1]{\mynote{Franz}{#1}{olive}}

\newcommand{\myparagraph}[1]{\smallskip{}\noindent{\textbf{{#1.}}}}

\newcommand{\mysubsubsection}[1]{\stepcounter{subsubsection}\smallskip{}\noindent{\textbf{\arabic{section}.\arabic{subsection}.\arabic{subsubsection} {#1}}}}

\newcommand*{\eg}{e.g.\@\xspace}
\newcommand*{\ie}{i.e.\@\xspace}
\newcommand{\sys}{\textsc{SinClave}\xspace}

\newacronym{ias}{IAS}{Intel Attestation Service}
\newacronym{sgx}{SGX}{Software Guard eXtensions}
\newacronym{tee}{TEE}{Trusted Execution Environment}
\newacronym{sdk}{SDK}{Software Development Kit}
\newacronym{cots}{COTS}{Components-off-the-Shelf}
\newacronym{os}{OS}{Operating System}
\newacronym{rop}{ROP}{Return-Oriented Programming}
\newacronym{afl}{AFL}{American Fuzzy Lop}
\newacronym{sigstruct}{Sig\-Struct}{Enclave Signature Structure}
\newacronym{tcb}{TCB}{Trusted Computing Base}
\newacronym{secs}{SECS}{SGX Enclave Control Structure}
\newacronym{avx}{AVX}{Advanced Vector Extensions}
\newacronym{cet}{CET}{Control-flow Enforcement Technology}
\newacronym{einittoken}{EINITTOKEN}{EINIT Token}
\newacronym{flc}{FLC}{Flexible Launch Control}
\newacronym{faas}{FaaS}{Function as a Service}
\newacronym{iaas}{IaaS}{Infrastructure as a Service}
\newacronym{saas}{SaaS}{Software as a Service}
\newacronym{sfi}{SFI}{Software Fault Isolation}
\newacronym{cas}{CAS}{Configuration and Attestation Service}

\begin{document}
\pagestyle{fancy}


\fancyhead{
        \vspace{-40pt}
        \begin{tikzpicture}
        \node[align=center] () at (0,0) {
                \begin{tcolorbox}[colback=yellow!40,
                colframe=white,
                width=\textwidth,
                boxrule=0mm,
                sharp corners]
                \centering
                CC-BY 4.0. This is the author's version of the work. The definitive version is published in the proceedings of the
                24th ACM/IFIP International Middleware Conference (Middleware 2023).
                \end{tcolorbox}
        };
        \end{tikzpicture}
}
\fancyfoot{}
\cfoot{\thepage}

\renewcommand{\headrulewidth}{0pt}


\title[SinClave]{SinClave: Hardware-assisted Singletons for TEEs}

\author{Franz Gregor}
\affiliation{%
  \institution{Scontain GmbH}
  \country{Germany}
}
\author{Robert Krahn}
\affiliation{%
  \institution{TU Dresden}
  \country{Germany}
}
\author{Do Le Quoc}\authornote{Do Le Quoc performed this work at TU Dresden.}
\affiliation{%
  \institution{Huawei Research}
  \country{Germany}
}
\author{Christof Fetzer}
\affiliation{%
  \institution{TU Dresden}
  \country{Germany}
}

\renewcommand{\shortauthors}{Gregor et al.}

\settopmatter{printfolios=true}
\begin{abstract}
For trusted execution environments (TEEs), remote attestation permits establishing trust in software executed on a remote host.
It requires that the measurement of a remote TEE is both complete and fresh:
We need to measure all aspects that might determine the behavior of an application, and this measurement has to be reasonably fresh.
Performing measurements only at the start of a TEE simplifies the attestation but enables ``reuse'' attacks of enclaves.
We demonstrate how to perform such reuse attacks for different TEE frameworks. We also show how to address this issue by enforcing freshness -- through the concept of a singleton enclave -- and completeness of the measurements. Completeness of measurements is not trivial since the secrets provisioned to an enclave and the content of the filesystem can both affect the behavior of the software, \ie{}, can be used to mount reuse attacks.
We present mechanisms to include measurements of these two components in the remote attestation. Our evaluation based on real-world applications shows that our approach incurs only negligible overhead ranging from $1.03$\% to $13.2$\%.

\end{abstract}

\maketitle
\thispagestyle{fancy}

\section{Introduction}


\gls{tee} technologies, such as Intel \gls{sgx}, have gained much attention in the industry~\cite{AzureSGX, IBMCloudSGX} as well as in academia~\cite{arnautov2016scone,costan2016intel,tsai2017graphene,shen2020occlum}.
\glspl{tee} can protect cloud-native applications (code and data) not only at rest, in transit, but also during computation (in use) against powerful attackers with privileged access to the underlying system software (\eg{}, the operating system or the hypervisor), as well as, hardware.
To ensure the confidentiality and integrity of applications, \glspl{tee} execute their code and data inside an encrypted memory region called \emph{enclave}. 
Adversaries with privileged access cannot read or interfere with the memory region and only the processor can decrypt and execute the application inside an enclave.  

Various hardware-assisted \glspl{tee} exist today, \eg{}, Intel SGX~\cite{Anati2013a,McKeen2013}, ARM TrustZone~\cite{TrustZone},  AMD SEV SNP~\cite{AMD-SEV-SNP}, and Intel TDX~\cite{TDX}.
Out of these \glspl{tee}, Intel SGX has been widely adopted in practice~\cite{AzureSGX,IBMCloudSGX} since 2015.
It enables a general-purpose \emph{ring 3}-enabled TEE and yields a significantly smaller \gls{tcb} in comparison with other solutions where typically a whole \gls{os} is part of the \gls{tcb}.
Furthermore, running legacy applications without source code changes is supported by several SGX platforms~\cite{arnautov2016scone, priebe2019sgx, tsai2017graphene, occlum}.

Arguably, just executing legacy applications in Intel SGX environments is not enough to ensure strong security guarantees.
First, the legacy applications were developed for a weaker threat model that trusts the underlying system software.
Thus, the \gls{sgx} frameworks must employ additional protection mechanisms to protect the naive legacy application against the elevated threats in the \gls{sgx} threat model.
An example is the file system for which \gls{sgx} frameworks transparently encrypt file content before it is persisted by the \gls{os}.
The content's integrity is additionally verified when files are read.
Second, deployment of legacy applications becomes significantly more complicated in \glspl{tee} as configuration data (containing secrets) for these applications must be hidden from potentially malicious system software.

To handle these problems, \gls{sgx} frameworks, such as Gra\-phe\-ne-SGX~\cite{tsai2017graphene} or Occlum~\cite{occlum}, only provide the building blocks for users and require individual implementation of secure data provisioning.
This approach contradicts the common goal to \emph{support legacy applications} without changing their source code.
Other frameworks, \eg{}, SCONE~\cite{palaemon,arnautov2016scone} and SGX-LKL \cite{priebe2019sgx}, provide an integrated solution that 
transparently serves software configurations only after it was verified that the \gls{tee} is genuine and that an adversary did not alter the application.

The verification process is called \emph{Remote Attestation}~\cite{costan2016intel} and allows one to establish trust in the application running inside an enclave on a remote host.
Remote attestation is the first step one performs to ensure that an application on a remote host, runs the correct code inside a genuine hardware enclave and it was initialized correctly.
After that, users establish a secure connection to the enclave to transfer secrets and configuration to their application.

We show that current remote attestation mechanisms for \glspl{tee}, \eg{}, Intel SGX, struggle to provide the expected trust and present an attack that bypasses the remote attestation protocol.
We further establish that the measurement during attestation needs to be both \textit{complete}, \ie{}, measure all aspects that determine the behavior, and \textit{fresh}, \ie{}, the measurement reflects the current behavior.
During runtime, applications may load libraries or open shared volumes, which may change the behavior.
The measurement should reflect the dynamic behavior, and ensure the freshness (to protect against ``reuse'' attacks), of the legacy applications running inside enclaves.

Unfortunately, performing measurements with current attestation mechanisms of TEEs only ensures an application's integrity at the starting of an enclave.
Thus, it's vulnerable to ``reuse'' attacks, as we show in \S{}\ref{sec:attack}: one can use a previous measurement to impersonate a TEE with a different expected behavior.
The vulnerability is based on the fact that almost all applications dynamically load data, libraries, and configurations during runtime, which can change their behavior.
An attacker adds malicious code via libraries or manipulates the configuration loaded into enclaves or loads data that might trigger some application bug. 
They can then steal the secrets of an application by bypassing the remote attestation as sketched in the following example.

Imagine a service provider (i.e., cloud user) who wants to run a Python application securely in a \gls{tee} in the cloud, \eg{}, an AI application based on PyTorch or TensorFlow. 
The service provider triggers the application's cloud deployment and subsequently uses remote attestation to assure the \gls{tee} was constructed without manipulations before securely sending the application's secrets.
The attacker can observe and intercept this attestation process by starting the requested Python interpreter correctly in the \gls{tee} but configuring it such that it will load a malicious library to execute a \gls{tee} impersonator program. 
This impersonator program then will correctly engage in the remote attestation procedure with the service provider and obtain his secrets for the Python application.
The service provider cannot detect this manipulation since the attestation procedure only captures the \gls{tee}'s initial state which the attacker did not alter.

Vulnerabilities in a program's configuration handling code -- typically not considered a vector of attack -- can make any program exploitable to the ``reuse'' attack. 
Considering this attack, Haven~\cite{baumann2015shielding} proposed to incorporate configurations into the \gls{tee} such that the attestation would also include them.
However, this would also reveal the \gls{tee}'s configuration and, thus, its secrets.
An alternative solution integrates the service provider's cryptographic identity into the \gls{tee} and enforces that the application would only start with correctly signed configurations.
This approach decreases the system's usability as every \gls{tee} needs to be individualized, prohibiting established procedures for software distribution and deployment.

We overcome the portrayed issues by proposing a novel protection mechanism against ``reuse'' attacks on remote attestation.
Our solution does not break the usability of legacy applications, \eg{}, supporting secure dynamic configuration and secure mounting of volumes, and incurs only a negligible overhead. 
Our contributions are as follows: 
\begin{itemize}
    \item We provide a comprehensive description of remote attestation in TEEs, especially Intel SGX (\S{}\ref{subsec:intel-sgx-remote-attestation} and \S{}\ref{sec:attestation}).
    \item We illustrate a practical attack on \gls{tee} secret provisioning, subverting all \gls{tee} protection goals (\S{}\ref{sec:attack}).
    \item We highlight a previously neglected vector of attack for \gls{tee} software rendering the majority of systems vulnerable to the presented attack (\S{}\ref{sec:attack}).
    \item We show the novelty of our attack by exploiting state-of-the-art \gls{tee} runtime systems (\S{}\ref{sec:attack_overview}). 
    \item We implement and evaluate a new flexible secret provisioning mechanism preventing the presented attack while supporting changes of input files, arguments, environment files etc., as well as software updates (\S{}\ref{sec:protection}).
    \item We conduct micro- and macro-benchmarks to demonstrate that our protection mechanism incurs only negligible overhead and can be used in practice (\S{}\ref{sec:evaluation}). 
\end{itemize}

\section{Background}\label{sec:background}

\subsection{Trusted Execution Environments}
Trusted Execution Environments (TEEs)~\cite{enclaves1,enclaves2} such as Intel SGX~\cite{AzureSGX, sgx-explained}, ARM TrustZone~\cite{TrustZone}, AMD SEV SNP~\cite{AMD-SEV-SNP}, and Intel TDX~\cite{TDX}, support the design of complex, yet secure, applications.
A \gls{tee} is a tamper-resistant processing environment that guarantees the authenticity, integrity, and confidentiality of its executing code, data, and runtime states (e.g., CPU registers, memory, and sensitive I/O).
The content of \gls{tee} remains resistant to software attacks even from privileged code, as well as any physical attacks performed on the main memory of the system.
In addition, \glspl{tee} offer \emph{attestation} for proving their trustworthiness to third parties.
However, we demonstrate the vulnerability of remote attestation by demonstrating a serious ``reuse'' attack on the \gls{tee} of Intel \gls{sgx} (\S{}\ref{sec:attack}).
We also introduce a mechanism to protect against this attack in practice (see \S{}\ref{sec:protection}).

\subsection{Intel SGX Security Argument}
\label{subsec:intel-sgx-remote-attestation}

Before introducing the attack that bypasses remote attestation, we provide a detailed description of the technical building blocks that our work builds upon.
These are primarily the \gls{sgx} enclave creation, measurement, and attestation procedures.
Enclave measurement and attestation are implemented mostly in hardware, such that they cannot be modified by a user.
Therefore, we have to rely on existing functionality to establish a practical solution that is usable with current hardware.
In the following, we will explain the fundamentals of the security argument for attestation in \gls{sgx} and show how existing frameworks use \gls{sgx} attestation and the resulting ramifications.
We further introduce the concepts for validating that an enclave has not been tampered with before its initialization.
We omit aspects of \gls{sgx} that are superfluous for the understanding of this work and that have been sufficiently explained elsewhere~\cite{costan2016intel}.

\emph{Enclave verification} begins with the creation of the enclave itself:
Representations of each executed operation are hashed into a measurement value.
Enclave creation is finalized with enclave initialization during which signer-based verification properties are bound to the enclave.
Once initialized an enclave can engage in remote attestation
to convince the verifier of its integrity.

\begin{figure}[!t]
    \centering
    \includegraphics[scale=0.5]{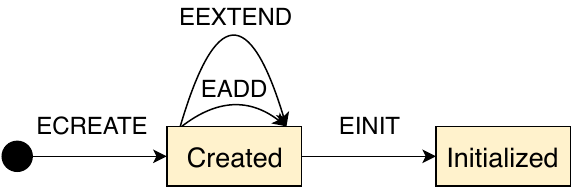}
    \caption{Enclave Creation Procedure.}\label{fig:enclave_creation}
    \Description{No description.}   
\end{figure}

\mysubsubsection{Enclave Creation}\label{sec:background:enclave_creation}

An \gls{sgx} enclave is created by a program referred to as the \texttt{starter}, which is not part of the \gls{tcb}~\cite{guide2020intel}.
The \texttt{starter} utilizes four \gls{sgx} instructions to create the enclave as shown in Figure~\ref{fig:enclave_creation}.
\texttt{ECREATE} creates an enclave, more precisely, its metadata \gls{secs} and, therefore, determines its maximal size and enclave attributes, such as 64-bit mode, debug mode, or whether it requires certain extended processor features such as \gls{avx} or \gls{cet}.
Once the enclave is created, \texttt{EADD} and \texttt{EEXTEND} instructions are used to add each page (4\thinspace{}kB) of code and data to the enclave's memory and extend its measurement respectively. 
Once the enclave's memory is set up, the enclave is initialized with \texttt{EINIT} preventing any further modifications of the enclave from the outside.
In the initialized state the enclave can be entered and assumes control over itself.

\begin{figure}[!t]
    \centering
    \includegraphics[width=0.88\linewidth{}]{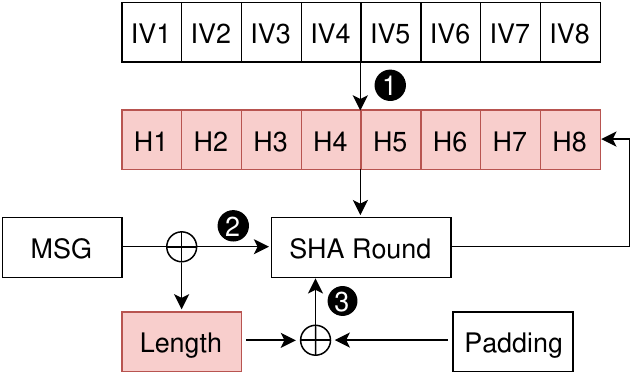}
    \caption{Schema of SHA256.}\label{fig:sha}
    \Description{No description.}
\end{figure}


Our solution tightly interacts with the \emph{hash algorithm} used for enclave measurement.
Hence, we provide the necessary detail about the utilized SHA-256 algorithm.
SHA-256~\cite{penard2008secure, pub2012secure} is one variant of the Secure Hash Algorithm 2 cryptographic hash functions.
The SHA hash functions are based on Merkle-Damg\r{a}rd construction~\cite{merkle1979secrecy} with Davies-Meyer compression functions.
SHA-256 maintains an internal state of 256-bits that are organized in 8 values (\(H1\) \dots \(H8\)) with 32-bit each that eventually becomes the hash value.
An arbitrary message, whose length is only limited by the hash internal length counter of 64-bit, is compressed into these 8 values.
Calculating a SHA-256 hash value can be summarized in three steps (see Figure~\ref{fig:sha}):

\emph{Initialization \raisebox{.5pt}{\textcircled{\raisebox{-.9pt} {1}}}}:
The hash state is initialized with initialization vector values according to the SHA-256 standard.

\emph{Updating \raisebox{.5pt}{\textcircled{\raisebox{-.9pt} {2}}}}:
Each 64-byte chunk of message data is processed iteratively with a procedure, we summarize as SHA Round and by updating the length counter.
The SHA Round consumes the incoming message chunk and intermingles it with the current hash state (\(H1\)\dots{}\(H8\)).
The output of the SHA Round becomes the new hash state.

\emph{Finalization \raisebox{.5pt}{\textcircled{\raisebox{-.9pt} {3}}}}:
After the entire message is consumed through iterative updating, the hash calculation is finalized yielding the hash value.
For this, the message length counter value is hashed into the hash state using the SHA Round procedure.
The 64-bit message length is padded according to a standardized padding scheme to fill the entire 64-byte input of the SHA Round.
This final result of the SHA Round calculation is the SHA-256 hash value of the input message.


Note that, due to the construction of the hash algorithm, for every \(64\) Bytes of compressed input an internal intermediary state is reached that can be described with \(256\) bit of internal hash state and \(64\) bit of already compressed input.
During an \gls{sgx} enclave measurement, each operation's representation is a multiple of \(64\) Bytes in length.
It follows that, after each enclave construction operation the enclave measurement hash calculation is in a state that can be described as said before.

\mysubsubsection{Enclave Initialization}\label{sec:background:enclave_initialization}

Before locking the enclave from further modifications, \eg{}, from system software, the enclave initialization verifies the \gls{sigstruct}.
The \gls{sigstruct} is created by the enclave signer, \eg{}, the software owner, and specifies the expected enclave measurement value (\texttt{MRENCLAVE}),
enclave attributes, the signer assigned product id, and a security version number.
The information is protected by an RSA signature created with the enclave signer's key.
Upon enclave initialization, the processor will ensure that the \gls{sigstruct} has not been manipulated and that the constructed enclave matches it.
To summarize the \gls{sigstruct} enables the signer to specify the \gls{sgx} configuration for a particular enclave and to specify access to specific \gls{sgx} sealing keys.

The \gls{sgx} \gls{sigstruct} can not prevent manipulations entirely.
As it is handled by system software, the adversary is free to modify it and subsequently sign it with their own key.\footnote{An adversary may execute an enclave with a configuration not intended by the signer to allow for exploitation.}
Therefore, the signer's identity (\texttt{MRSIGNER}) is incorporated during the attestation procedure (detailed next).

Another mechanism utilized during enclave initialization is the verification of the \gls{einittoken} data structure.
An \gls{einittoken} allows an enclave with a certain \texttt{MRENCLAVE} measurement, \texttt{MRSIGNER} identity, and enclave attributes to start.
It is generated by a dedicated system enclave, the \texttt{launch enclave}, and must be provided to the \texttt{EINIT} instruction initializing the enclave for it to succeed.
In the first implementation of \gls{sgx}, only enclaves signed by Intel were allowed to become launch enclaves.
The launch enclave from Intel generated \gls{einittoken}s for any enclave in debug mode.
However, enclaves in production mode must be signed with a signer's key that is authorized by Intel to obtain a \gls{einittoken} data structure.
Later Intel introduced \gls{flc} allowing the platform owner to determine which launch enclaves can run.
With \gls{flc} the default launch enclave allows any enclave to run in production mode.

\mysubsubsection{Attestation with SGX}\label{sec:background:attestation}

When an enclave is running (initialized and entered) it can request a report from the processor.
The report entails the enclave's \texttt{MRENCLAVE}, \texttt{MRSIGNER}, enclave attributes as well as information about the platform's processor.
With the help of a quoting enclave, the report can be transformed into a remotely verifiable quote.
Note that the report and quote show the enclaves' state at the point of initialization.
That is, any modifications afterward (by the enclave) are not reflected in the quote and are, thus, not observable by the remote verifier.

Intel has defined protocols for local and remote attestation to be used for verifying the genuineness of an SGX enclave and the software executed within the SGX enclave~\cite{Anati2013a}.
\gls{sgx}'s attestation scheme relies on a key generation facility and on a provisioning service.
During the manufacturing process, a ``Seal Key'' and a ``Provisioning Key'' are burned into the die of an SGX-enabled processor.
The provisioning secret is generated by Intel at the key generation facility and also stored by the provisioning service for later identification.
The seal key is generated by the processor and not known to Intel~\cite{sgx-explained}.


Intel distinguishes between local and remote attestation.
Both describe a process of an enclave proving its identity to a second dedicated enclave.
As a result, two enclaves on the same platform have established trust and can exchange information. 

During attestation a dedicated instruction (\emph{EREPORT}) defined by the SGX-SDK is used to create a report that cryptographically bundles, most importantly, a measurement of the enclave (\emph{MRENCLAVE}), a certificate (identity information) of the enclave's software (\emph{MRSIGNER}) and a hardware MAC to prove authenticity.
Other information such as the CPU's security version number is included as well but herein not further explained.
The attestation report can then be used to verify that it was produced by identified software isolated by genuine (Intel-) hardware.

Remote attestation is executed after the local attestation has been executed successfully and a local quoting enclave has established trust in the to-be-attested enclave.
Through remote attestation, a third party (challenger) can remotely establish trust in an SGX-enclave and ensure the genuineness of the executed soft- and hardware.
Section~\ref{sec:attestation} explains attestation in more detail.

\subsection{System \& Threat Model}\label{sec:threat_model}

All systems building upon \gls{sgx} must rely on its hardware-assisted attestation mechanism to establish trust.
In \gls{sgx}' threat model the processor is the only trusted component.
In this section, we assume an attacker who is not able to violate the security guarantees of \gls{sgx}.
However, attackers can create enclaves in the same environment as the user to impersonate the user's enclave and create attestation reports with arbitrary \texttt{reportdata} content.

Our system model has the following assumptions:
\begin{itemize}
    \item The software is developed and distributed by a trusted software developer.
    \item A trusted signer prepares the software for execution inside of \gls{sgx}. This is done with the help of a trusted \gls{sgx} framework that enables the execution of legacy software inside of \gls{sgx}. In particular, the signer creates a valid \gls{sigstruct}. 
    \item The user customizes the \gls{sgx}-enabled software by configuring it. The configuration (files, environment variables, program arguments, etc) is secured through trusted means provided by the \gls{sgx} framework.
    \item Developer, signer and user and their systems are benign and have means to securely exchange information about the integrity of the software.
    \item The customized software is transmitted and executed in an \gls{sgx}-enabled environment under control of the adversary.
    \item The utilized \gls{sgx}-environment delivers its security goals. For example, the remote attestation mechanism and infrastructure are trusted and we assume \gls{sgx} processor implementation is not exploitable.
    \item The attestation protocols are known to the adversary, \ie{}, no security through obstruction.
\end{itemize}

Note that the developer, signer and/or user roles might be shared by the same entity.
SGX cannot protect secrets at rest for previously unknown platforms.
Therefore, secrets can not be part of an enclave but must be provisioned after the enclave has been found trustworthy via remote attestation.

In our system model, the user wants to run a native application in an enclave in a public cloud.
The enclave needs a configuration to run and secrets to, \eg{}, authenticate to other services or decrypt sealed file system content.
The user utilizes \gls{sgx} remote attestation to build trust into the remote enclave before sending its configuration via a secure channel.
The secure channel is bound to the enclave via attestation.
The attestation and configuration tasks are delegated to a trusted verifier service on behalf of the user.

Note that our system model is closely modeled after what we see in existing systems:
For example, SGX-LKL 
enclaves load an encrypted disk image and set program arguments and environment variables according to the configuration~\cite{priebe2019sgx}.
The trusted verifier service is the \texttt{sgx-lkl-ctl} command line tool that attests the enclave and delivers the configuration.
The secure channel is based on wireguard and it is bound to the enclave by adding the enclave's public wireguard key in its reportdata field.
In a SCONE, program arguments, environment variables, and session secrets are retrieved from the configuration and encrypted file system content is opened before the internal program is loaded.
The SCONE \gls{cas}~\cite{palaemon} is the trusted verifier service to which the enclave connects and from which it retrieves its configuration through encrypted channel.
Recently, remote attestation was added to Graphene-SGX~\cite{tsai2017graphene} as a shared library that uses RA-TLS as a secure channel bound to the enclave.

\section{Remote Attestation Bypass Attack}\label{sec:attack}


        
        

In this section, we describe the remote attestation bypass attack in which the adversary uses an enclave trusted by the user to gain the user's secrets.
First, we provide an overview of remote attestation in \glspl{tee}, then we show the vulnerability of the general remote attestation protocol provided in \glspl{tee}.
Afterward, we describe the attack against remote attestation in Intel \gls{sgx}.

\begin{figure*}[!t]
	\centering
	\begin{minipage}{.48\linewidth}
	\vspace{4pt}
	\includegraphics[width=.97\textwidth]{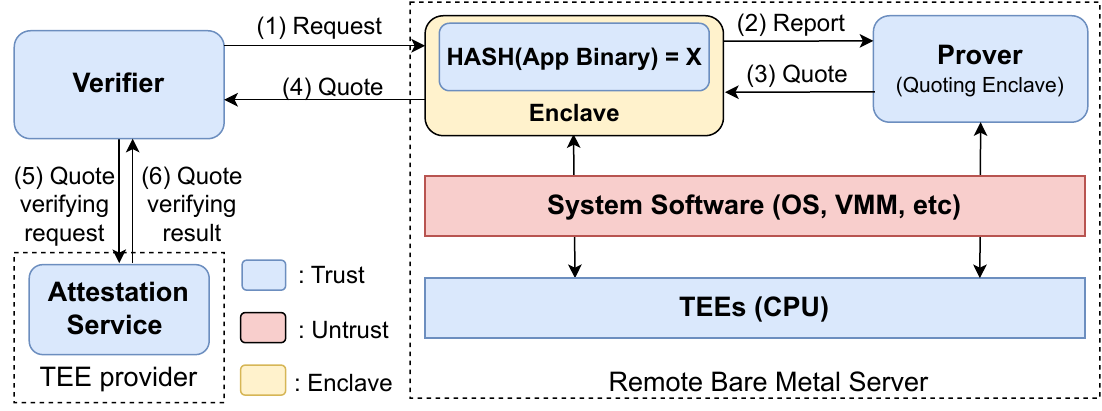}
	\vspace{13pt}
	\caption{General remote attestation protocol in \glspl{tee}.}\label{fig:remote-attestation-overview}
	\Description{No description.}
	\end{minipage}
    \hspace{15pt}
	\begin{minipage}{.48\linewidth}
    \includegraphics[width=.97\textwidth]{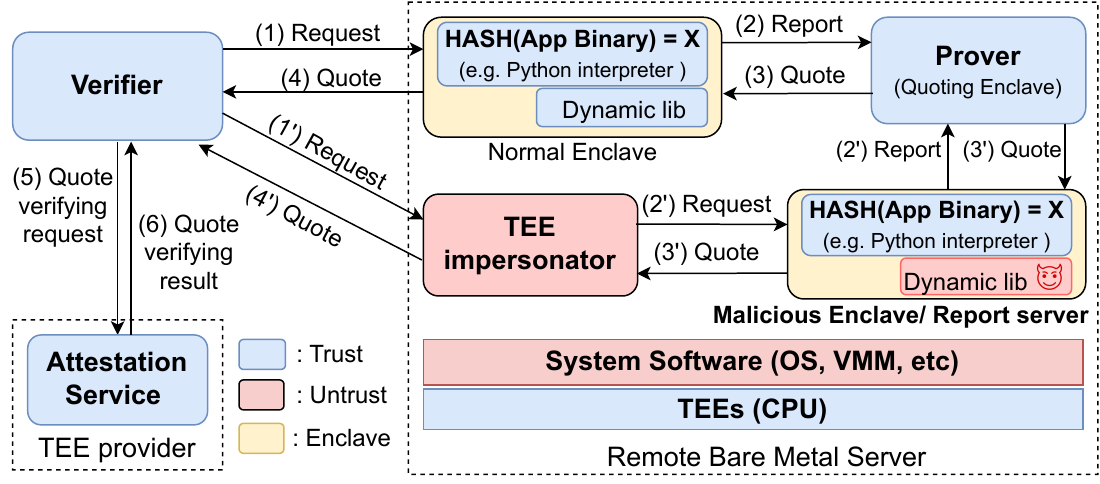}
	\Description{Overview of remote attestation attack: If the application dynamically loads libraries and its configuration during runtime, an attacker can load compromised libraries or manipulate the configuration of enclaves.}
	\caption{The overview of remote attestation attack.}\label{fig:remote-attestation-attack-overview}
	\end{minipage}%
\end{figure*}



\subsection{Attestation Overview}\label{sec:attestation}


Remote attestation provides the ability for a verifier to assert a certain state of an application is running in a \gls{tee} (see Figure~\ref{fig:remote-attestation-overview}).
On the device, there is a prover entity (\emph{Quoting Enclave} in \gls{sgx}) able to produce the evidence for the state of the application before running it; and securely send it to the verifier.
The attestation process consists of two major parts: \emph{(i)} Characterization of the application's current state (the measurement process to create the evidence, called \emph{quote} in \gls{sgx}); and \emph{(ii)} Secure transfer of the evidence to the verifier.

The measurement is used to prove the application is running in a correct state (\eg{}, correct code) inside enclaves of \glspl{tee}.
It can be a cryptographic hash over the binary of an application (see \S{}\ref{sec:background}) or could also contain other target information for the attestation.
The measurement is then signed using a hardware-protected certified key unique to the device to create the evidence or quote that can be verified by the verifier. 
The certified key ensures the \emph{authenticity} of the remote attestation protocol.
To ensure the \emph{freshness} of the evidence, the verifier includes a \emph{nonce} into the challenge in its request to the remote enclave.

Figure~\ref{fig:remote-attestation-overview} summarizes the general remote attestation workflow in \glspl{tee}.
Before launching an application in an enclave deployed on the remote device/host, the measurement process creates the attestation evidence.
When the verifier wants to verify if the remote application has an expected state, it sends a request with a nonce to the remote enclave (step (1)).
The enclave combines the nonce with the measurement, signs it to create a report using attestation keys (stored in CPU hardware), and sends it to the prover (step (2)).
The prover verifies the report, signs it to create the evidence (\ie{}, quote), and sends it back to the verifier (step (3) and (4)).
The verifier checks the signature and certificate with the help of the trusted third-party, \eg{}, the attestation service of the \gls{tee} provider,
and checks that the measurement indicates the start of the expected enclave (steps (5) and (6)).
Note that in other TEEs, e.g., AMD SEV SNP, the enclave is a secure virtual machine (VM) and the remote attestation and quoting are performed by the AMD Secure (Co-)Processor.

\vspace{-2mm}\subsection{Remote Attestation Bypass Attack Overview}\label{sec:attack_overview}

A significant security issue in remote attestation of current TEEs is
that the measurement process is only conducted before running
the application and cannot be performed during runtime. 
If the application dynamically loads libraries and its configuration during runtime, an attacker may load compromised libraries or manip-
ulated configurations into enclaves (see Figure~\ref{fig:remote-attestation-attack-overview}).

The attacker can create a genuine evidence (quote) using the malicious enclave since it has correct measurement and send this evidence to the verifier to bypass the remote attestation (steps (2), (3), and (4)). 
After bypassing the remote attestation, the attacker can steal secrets (details in \S{}\ref{subsec:sgx-attack}). 
Applications with the pre-compiled binaries (e.g., Python) are vulnerable to this attack since their configuration mainly determines their behavior.

In this work, we imagine a user who wants to run their legacy application in a potentially adversarial cloud environment. 
The user protects the application’s code and data during runtime using
an \gls{sgx} framework. 
At runtime, the application’s behavior significantly depends on data the application consumes (mainly from the file system), e.g., configuration files or mounted volumes. 
Remote attestation is subsequently used to establish that the application has not been modified by an adversary prior to execution.

The adversary needs two components for the attack: an \textit{TEE impersonator} and a \textit{report server}.
The \gls{tee} impersonator implements the attestation protocol utilized by the verifier, while the report server generates \gls{sgx} reports with arbitrary, adversary chosen \texttt{reportdata} fields.
Note that the \textit{TEE impersonator} is not required to be executed inside of an enclave.
The \textit{report server}, on the other hand, must be executed inside of an enclave, fulfilling expectations by the verifier and allowing it to generate \gls{sgx} reports.
Next, we describe how the adversary can build a \textit{report server} from an enclave without this being reflected in the enclave's measurement.

\myparagraph{Creating a Report Server}\label{sec:report_oracle}
A report server is an enclave that produces \gls{sgx} reports with arbitrary report data upon the adversary's instruction.
The concrete steps and complexity of creating a report server from an arbitrary enclave varies but three operations must be implemented:

\begin{itemize}
    \item The desired report data and target information must be copied into enclave memory.
    \item The \gls{sgx} report generation must be invocable.
    \item The created report must be delivered to a verifier.
\end{itemize}

Network connections, file system interfaces, the terminal, or reading of environment variables can be utilized for communication.
Arguably any enclave must provide some form of input and output to do useful work.
We also assume that any enclave of interest would allow for \gls{sgx} report generation given that \gls{sgx} attestation is to be used to securely provision configurations to the enclave and that remote attestation can be used throughout the lifetime of an enclave.
For example, attestation is used to convince users of a service of its trustworthiness or sign data produced by the enclave.
Many \gls{sgx} frameworks voluntarily expose attestation features including setting arbitrary \texttt{reportdata}:
SCONE~\cite{arnautov2016scone} exposes report generation via C functions to user code.
Occlum~\cite{shen2020occlum} provides special \texttt{ioctl} system calls.
Gramine~\cite{Gramine} (initially Graphene-SGX~\cite{tsai2017graphene}) exposes the report generation via pseudo-files in \texttt{/dev/attestation}.
Hence, we assume that the three necessary operations are present in any enclave an adversary would be interested in.

With these building blocks available, the adversary needs to gain control over the enclave such that it uses these operations according to the adversary's desire.
Such control is obtained by exploiting vulnerabilities in the enclave's code and, thus, many frameworks use memory-safe languages to mitigate such risks.
However, we found a significantly simpler approach: through configuration.


\myparagraph{Creating a Report Server by Configuration}
Our threat model allows adversaries to use significantly wider attack vectors than in a more traditional model.
In particular, they can start and configure the user's enclave arbitrarily often.
They can also run their own verification and configuration component that delivers the desired configuration to the enclave.
In summary, if not prohibited by the enclave itself, the adversary can run the user enclave with an arbitrary configuration.
%

If the software's configuration that is not reflected in the enclave measurement (\S{}~\ref{subsec:sgx-attack}) dictates the to-be-loaded program, the adversary can simply start a report server implementation.
In other cases, the user application might be substantially configurable.
This is mostly always the case for language interpreters, for languages such as Java, Python, JavaScript, Ruby, etc.
In some frameworks, these interpreters are already the entire enclave while the actual program is loaded from the file system at runtime.
For example, when Java applications, such as Wildfly/JBoss or Tomcat, are executed with the exact same interpreter and, thus, the enclave is used.
Thus, the adversary can configure the interpreter enclave to load their report server implementation in the necessary language, for example, Java without the attestation report reflecting this.

Alternatively, the adversary can use any program that loads dynamic code during runtime. 
Dynamic loading is broadly used to load optional features into a program. 
For example, Apache httpd loads only the configured modules, or NGINX’s dynamic module extends its features using dynamically loaded code.
Since the dynamic code is only loaded during runtime, it is typically not reflected in the enclave measurement and, thus, manipulating it is not revealed through attestation. 
Therefore, the adversary can patch existing functions such that they execute the desired operations. 
This code could read the desired reportdata from a file, query the SGX report, and write it into another file implementing a report server. 
With this, the adversary could trigger the code by querying a site with the appropriate compression algorithm requested.

A straightforward way to handle the attacks is using static linking for applications. 
However, it also comes with the cost of increased binary size which makes booting time slow, compatibility issues, performance overhead, and resource redundancy. 
For interpreter languages, e.g., Python they need to pack dependencies with interpreter engines which is cumbersome. 
It requires heavy engineering, especially with complex libraries such as TensorFlow. 

Besides SGX, other TEEs, such as AMD SEV and Intel TDX, have a similar issue since these TEEs only perform measurements and ensure the integrity of confidential VMs after the booting process. During computation, these TEEs do not perform any measurements if shared data or libraries from the host are loaded to the VMs; it requires carefully checking or adding their measurement during the boot process. If there is no shared data or libraries between VMs or with the host, an attacker can just boot the VM from a victim and perform, e.g., side-channel attacks on the VM in a lab environment.

\subsection{Attack Examples}\label{subsec:sgx-attack}
In this section, we show the practicability of the previously described attack by describing our attack against state-of-the-art \gls{sgx} frameworks.
Compared to previous work~\cite{van2020sgaxe}, our approach does not attack the sealing mechanism of \glspl{tee}.
We show our attack against applications running in SCONE as well as in SGX-LKL which support transparent remote attestation for users.

\mysubsubsection{Attacking Python and NodeJS in SCONE}

We assume the Python application's files are protected with SCO\-NE's file system encryption.
The application's configuration, including the Python interpreter's expected enclave measurement and signer identity and the file system encryption key, are stored in SCONE's \gls{cas}~\cite{palaemon} the secrets management service (running in an enclave itself).
SCONE enclaves connect to \gls{cas} during startup to obtain their configurations and transparently provide it to the executed application.
For this, \gls{cas} ensures that the enclaves match the user-provided expectations, utilizing \gls{sgx} attestation.
That is, in SCONE, \gls{cas} functions as the attestation verifier.

\myparagraph{TEE Impersonator}
For building a \gls{tee} impersonator for SCONE we adapted the existing CAS-client implementation.
The \gls{sgx} report retrieval mechanism of the client is configurable through a callback mechanism.
Hence, we only needed to implement a client for the SCONE report server implementation (discussed below) and configure the CAS client to use it.
Together with the necessary argument parsing the \gls{tee} impersonator implementation consists only of \(75\) lines of code.

\myparagraph{Report Server}
In SCONE, a program's file system is not reflected in the enclave measurement value.
Any Python program utilizing the same Python interpreter in SCONE uses an identical enclave and can not be distinguished by the verifier.
The verification of the actual Python (byte-) code that is stored in the encrypted file system is delegated to the runtime.
With the configuration, the runtime obtains the expected state of the file system and uses this information to ensure files have not been tampered with during execution.
This delegation is precisely the exploitable culprit as the passed configuration is not verifiable through attestation.
Based on this, an adversary can invoke the user's Python enclave and configure it to run a report server program instead of the user's application (after attestation).

For the attack, we implemented a \textit{report server} with a simple Python socket server that receives the information necessary to query the report, in particular the \texttt{reportdata}, via the network.
The \gls{sgx} report is obtained through a custom Python C module in which we use the \texttt{EREPORT} \gls{sgx} hardware instruction in assembly.
The resulting report is sent via the same network connection towards the \gls{tee} impersonator.
Our attack requires dynamic loading of code at runtime as required by the Python interpreter.
The entire report server implementation consists of \(33\) lines of Python code and \(76\) lines of C code.

For the attack on NodeJS, we implemented a separate \textit{report server} as the interpreter binaries.
NodeJS allows the execution of native (C) code through C++ addons that are loaded dynamically at runtime.
We implemented the server loop in \(22\) lines of JavaScript and the addons executing \texttt{EREPORT} required \(67\) lines of C++ code.

\myparagraph{Attack Procedure}
For the attack, the adversary first starts the \textit{report server} built from the enclave trusted by the user.
Then the adversary invokes the \gls{tee} impersonator configuring it with the user's configuration ID as well as the addresses of the necessary services and waits for the \gls{tee} impersonator to yield the user's configuration.

The \gls{tee} impersonator connects to \gls{cas} and requests the user enclave's configuration.
The connection is established via a secure TLS channel whose certificate key's are used to ensure the authenticity of the connection, \ie{}, ensuring the channels integrity but also that the channel is terminated in the expected enclave.
In response \gls{cas} will demand a suitable remote attestation quote (SCONE, EPID, or DCAP quotes are supported).
Depending on the system capabilities an appropriate mechanism is chosen by the impersonator and the necessary \gls{sgx} report is queried from the report server.
The report server is instructed to incorporate the \gls{tee} impersonator's certificate key into the report's \texttt{reportdata} field undermining the channel's authenticity.
The verifier (\gls{cas}) is unable to detect the manipulation because the quote is valid, it shows the expected enclave is running and the TLS channel utilizes the expected certificate whose key is incorporated into the report.
Thus, it grants access and delivers the requested configuration.

\mysubsubsection{Attacking Python in SGX-LKL}\label{sec:attack_lkl}

In SGX-LKL, the enclave starts a simple attestation and configuration service when invoked with \texttt{sgx-lkl-run}.
The user then connects to that service using the \texttt{sgx-lkl-ctl} tool and attests the enclave.
For this, the tool shows the \gls{sgx} report/quote details on the screen.
After verification of the shown details, the user invokes \texttt{sgx-lkl-ctl} again to transmit the enclave configuration, which contains the disk encryption key.
SGX-LKL uses encrypted disk images to load code into the enclave.
Having received the configuration, the enclave runtime accesses the disk and invokes the user application.

SGX-LKL implements additional restrictions to ensure the integrity of the attestation procedure.
For example, the runtime enforces that attestation and configuration is only done once.
The integrity of the attestation and configuration procedure is enforced by a public wireguard key that is provided to the enclave during invocation.
That is, the enclave will only interact with attestation and configuration requests originating from the benign user.
Furthermore, we assume the system is configured to run in release mode, which adds additional restrictions enforced by SGX-LKL:
The disk image (program) must be encrypted, and the SGX-LKL will not start the user program without attestation.
We will show beneath that only this last restriction influences the attack but does not mitigate it.

\myparagraph{TEE Impersonator}
SGX-LKL's attestation and configuration interface is implemented in C and uses \emph{protobuf} for message framing.
We implemented the SGX-LKL impersonator in Python.
It opens the attestation and configuration ports and waits for the user to connect.
SGX-LKL utilizes Wireguard (VPN) to establish secure channels.
We instruct the host \gls{os} to provide the VPN expected by the client instead of handling these details in the impersonator.
Upon start, the impersonator is configured with the nonce expected by the user and the public wireguard key the adversary configured the VPN with.
Our implementation consists of \(236\) lines of code and about \(2000\) lines of generated \emph{protobuf} Python code.

\myparagraph{Report Server}
The fact that SGX-LKL uses encrypted disk images and that the enclave is attested before the user code can be loaded -- since it is encrypted -- hints that attestation in SGX-LKL ensures the integrity of the framework but not the user application.
User application integrity verification is implemented by the runtime by verifying the disk image's integrity.
In consequence, two different programs running in SGX-LKL will, from \gls{sgx} attestation perspective, be the same as long as they use the same version of SGX-LKL.
Consequently, the verifier can not distinguish them posing little restrictions for our report server implementation.
We implemented it in \(152\) lines of C code.

\myparagraph{Attack Procedure}
The attack on SGX-LKL requires the adversary to interfere with the commands sent by the user.
The adversary is in control over the machine on which the user wants to execute the enclave which allows them to execute all necessary actions.
Before the actual attack, the adversary prepares an encrypted disk image containing the report server.
When the user attempts to start their enclave with \texttt{sgx-lkl-run},
she provides their public wireguard key and a nonce (and the desired disk image) to the command to ensure the enclave can only be controlled by them.
The invocation command is intercepted by the adversary and they start the report server's disk image, however, with the adversaries public wireguard key using the \texttt{sgx-lkl-run} command.
This enclave is then attested and configured by the adversary to execute the report server.
At the same time, the adversary starts the SGX-LKL impersonator and configures it to put a public wireguard key controlled by the adversary into the \gls{sgx} report.
Incoming requests to the user's enclave are forwarded the configuration interface of the SGX-LKL impersonator using the same wireguard guard key.
Afterward, the adversary waits for the attestation request from the user that is forwarded to the SGX-LKL impersonator.
It queries the report server for the necessary \gls{sgx} report containing the adversary's wireguard key in the \texttt{reportdata} field.
The report is sent back to the user for inspection.
At this point the user has no chance to detect the manipulation, seeing only a valid \gls{sgx} report for the expected SGX-LKL enclave.
Thus, they decide to trust it and use the embedded wireguard key to establish a secure connection to send the configuration over.
The wireguard key is embedded in the report
and the connection is established with the wireguard setup in the malicious host \gls{os} responded to by the impersonator, revealing the configuration to the adversary.



\section{Attestation Protection Mechanism}\label{sec:protection}

The remote attestation bypass attack uses fundamental technical properties built into \gls{sgx} rendering a verifier unable to distinguish between a newly started enclave without a configuration and an enclave pre-configured by an adversary.
In the following, we detail our protection mechanism against the remote attestation bypass attack.


\subsection{Design Space}\label{subse:protection-requirements}

Different approaches for protection mechanisms are possible due to the complex design space of remote attestation and the software stack involved.
However, we adhere to two common characteristics of modern cloud-native applications:
An enclave's configuration might not be known beforehand, \ie{}, program arguments of spawned sub-processes might be computed dynamically by the parent process and files can be used to transfer data between processes.
Hence, the protection must be \emph{flexible} to adapt to the dynamic behavior of applications and the protection mechanism must be practical, \ie{}, usable with existing hardware.


\subsection{General Solution Principle}

We prevent the shown exploitation by enabling the verifier to establish that an enclave is fresh and complete, \ie{}, it cannot have received a malicious configuration beforehand.
One possibility to prevent the attack would be to forestall the adversary configuring the enclave into a report server.
However, preventing the adversary from freely configuring the enclave would also contradict our \emph{flexibility} requirement.
Next, we discuss solutions with specific shortcomings before presenting our solution -- \emph{Singleton Enclaves} (\S{}\ref{sec:sin-enclave}).

\subsection{Naive Approaches}




\myparagraph{Restricted Runtime Operations}
A naive approach implemented in SGX-LKL allows the enclave to run only if it is attested and its configuration is provisioned via a secure channel that must be terminated by the trusted verifier.
Still, we can exploit it (as shown before) because the trusted verifier must also be configured in the enclave, and this configuration is not reflected in the attestation report and, thus, can not be verified.
Therefore, the adversary can start an enclave, configure it to trust the adversary, and start a report server.
This report server can then be used to fool the user by pretending to be a fresh SGX-LKL enclave since its \gls{sgx} report neither reflects that it was configured already nor that its actually SGX-LKL runtime was configured to trust the adversary.

\myparagraph{Using SigStruct}
As an alternative to the previous approach, the \gls{sgx} \gls{sigstruct} data structure could store additional information without the need for adaption of the software.
However, there are only very limited fields/space that can be freely controlled -- a couple of bytes.
Unfortunately, this is not enough to store a cryptographic identity but merely a config identifier. 
Using config identifiers is insufficient as their enforcement would have to be implemented in the verifier that is suspectable to manipulation by the adversary.
Imagine a \gls{sigstruct} stating that the enclave only accepts configurations with id A.
The adversary could simply accept that \gls{sigstruct} and ignore the restriction.
Alternatively, a hash of an entire config would be equivalent to the comprehensive enclave measurement solution but unfortunately would be too large for the data structure.

The \emph{MRSIGNER} field could also be used for the protection:
The enclave could only accept configuration that comes from a specific signer identity.
This approach would work, however, it requires the users have to sign all individual enclaves themselves. 
We regard this approach as impractical for cloud-native applications.

\fg{We could improve the argument by saying that a singleton enclave reduces the attack surface -- There is a python that only accepts configurations from B, but one python program has a bug and enables the adversary to build a report oracle -- no python program of B is secure anymore.
The \gls{sigstruct} offers fields that allow custom values that are signed by the signer key.
One can use these fields to implement singleton enclaves based on the \gls{sigstruct} information or even use unique signer keys.
}

\myparagraph{Using reportdata to show state}
As an alternative solution, one could include a config identifier or similar that proves the enclave was not configured already into the report data field of the report.
After all, the report must be created from the code that was measured in it and thus only that code could say it has not been configured at runtime.
This approach has been supported by SCONE and SGX-LKL but using the report data field for such purposes is not secure in our threat model.
After an adversary manages to create a report oracle, she also controls the report data.
Consequentially, only fields set by the hardware can be trusted against a powerful adversary.

\subsection{Singleton Enclaves}\label{sec:sin-enclave}
\begin{figure}[!t]
    \centering
    \includegraphics[width=.71\linewidth]{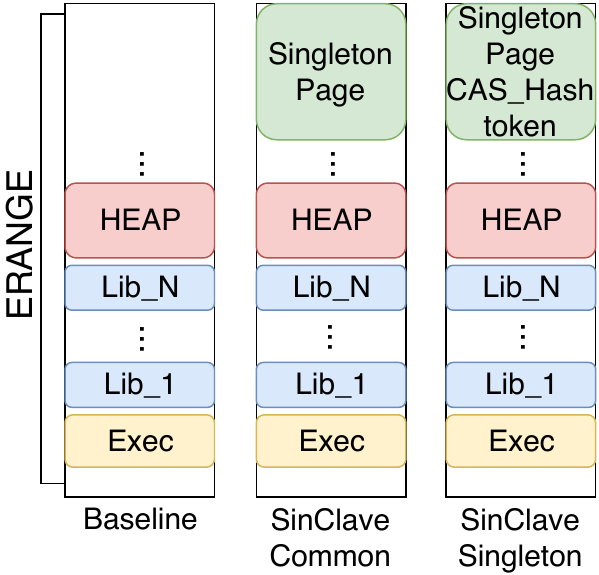}
    \caption{Enclave Memory Layout in \sys{}}\label{fig:solution:enclave_memory_layout}
    \Description{The Figure shows three memory layouts for enclaves. It is presented how \sys{} adds a single page to the enclave memory.}
\end{figure}

Our proposed protection mechanism is based on the following observations.
The attestation token is a mechanism that individualizes an enclave's measurement value (\texttt{MRENCLAVE}) in a verifiable manner such that re-attestation can be detected.
This individualization happens at enclave construction time and is thus compatible with the binary distribution of software.

The basic idea behind our solution is that system software adds an additional page, the \emph{instance page} (see Figure~\ref{fig:solution:enclave_memory_layout}), to the enclave dynamically during enclave construction.
This page, in particular, contains an attestation token, and the verifier's cryptographic identity.
The attestation token is unique and generated by the verifier in a previous step.

Once the enclave is started it directly engages in attestation to obtain its configuration from the verifier.
In the request, the enclave sends its \gls{sgx} quote and the attestation token to the verifier via a secure channel.
The verifier calculates the expected \texttt{MRENCLAVE} based on the expected enclave, the attestation token, and its own cryptographic identity and can thus, despite individualization, verify the enclave's integrity.

The enclave utilizes the verifier identity that is part of the \emph{instance page} to ensure it only obtains configuration for the intended verifier.
This prevents an adversary from supplying its malicious configuration and the verifier can utilize the attestation token to keep track of attested enclaves ensuring each enclave is attested exactly once.
In the following, we discuss the most important aspects of our approach.

\myparagraph{Verifiable Enclave Extension}
Adding an additional page to the enclave must necessarily yield the desired different enclave measurement.
On the other hand, the verifier must be able to ensure the intended enclave was extended correctly, \ie{}, it was neither manipulated beforehand nor incorrectly extended.
Adding an additional page to the end of an enclave requires the additional execution of one \texttt{EADD} and four \texttt{EEXTEND} \gls{sgx} instructions.
From the perspective of the enclave measurement SHA-256 calculation, adding an additional page to an enclave constitutes four additional update calls.
This allows the verifier to pre-calculated the expected enclave measurement:
The SHA-256 calculation's internal state needs to be instantiated just before the initial (not yet individualized) enclave was finalized with \texttt{EINIT} and updated according to the expected content and position of the \emph{instance page}.
Position and content of the \textit{instance page}, as well as, the necessary SHA-256 calculation state can be computed by the verifier given the enclave.\rk{Do we explain how?}

\myparagraph{On-Demand SigStruct Creation}\label{sec:sigstruct}
The \gls{sigstruct} is necessary to initialize an enclave (\S{}~\ref{sec:background:enclave_initialization}) but is only valid for a specific enclave measurement value (\texttt{MRENCLAVE}).
Since the attestation token individualizes enclave's \texttt{MRENCLAVE}, a suitable \gls{sigstruct} must be computed for each individualized enclave before it can be initialized.
Computing the \gls{sigstruct} requires knowledge of the enclave signer's private key.
This key must stay private: 
(a) If the adversary knows the key, we could only use the enclave measurement for verification.
(b) Any enclave with the signer identity A can derive signer identity A's \gls{sgx} sealing keys.
If the adversary could sign his own enclaves, the \gls{sgx} sealing keys would be compromised.
(c) In SGX v1 only (signer) identities whitelisted by Intel can run enclaves in production mode.
Compromised keys would be blacklisted.

In our protection mechanism, we distinguish \emph{singleton enclave} and \emph{common enclave}.
Both have the same base enclave hash -- the same software/code.
However, the singleton enclave must be attested and have a unique \emph{MRENCLAVE} while the common enclave can be started arbitrarily many times.
The \emph{instance page} of the common enclave is zeroed such that the runtime can decide whether it requires attestation or not.

When querying the verifier for the attestation token, the common enclave's \gls{sigstruct} is also sent to the verifier.
The verifier ensures it matches the base enclave hash (if instantiated for the common enclave) and if so finalizes the base enclave hash according to the attestation token it just chose.
The verifier uses that value to create the (on-demand) \gls{sigstruct} for the particular singleton enclave he expects to be started according to the base enclave hash, the token and his certificate.
The \gls{sigstruct} is equivalent to the common enclave's \gls{sigstruct} except that the enclave measurement and the signature is exchanged.
That \gls{sigstruct} is sent to the starter who will initialize the enclave with the instance/singleton page and the received on-demand \gls{sigstruct}.
This ensures that the signer key never leaves the trusted verifier and only enclaves the signer previously signed already (he trusted) will be attested.
In consequence, our protection satisfies the requirements described in \S{}\ref{subse:protection-requirements}.

Note that the goal of the proposed protection mechanism is to allow users to create a singleton SGX enclave that can be extended to a singleton confidential VM. In the singleton confidential VM, the measurement process during booting time includes the user identity and security for that VM. In other words, an attacker cannot just boot a VM of the user since he/she cannot generate the unique measurement of the VM. Only after verifying that the VM is a unique single instance, secrets are injected into the VM.

\vspace{-3mm}
\section{Evaluation}\label{sec:evaluation}

This section presents the experimental evaluation of \sys{} using micro- and macro-benchmarks.
In the evaluation, we focus on the system behavior expected to be influenced by our contributions.
These are enclave signing, enclave construction and attestation.
During the signing \sys{}'s interruptible SHA-256 implementation is used.
We expect it to impose a lightweight overhead since it is implemented using only Rust, while state-of-the-art SHA-256 implementations use optimized assembly code and hardware acceleration~\cite{gulley2013intel}.
An enclave based on \sys{} must query the verifier before the singleton enclave can be initialized.
This communication is necessary to get the attestation token and the on-demand \gls{sigstruct} and will slow down the overall application start.
Finally, the verification with \sys{} requires additional checks, which we expect to have only negligible influence.
After describing our evaluation settings (\S\ref{sec:eval:hardware-setup}), we discuss our results for the micro- (\S\ref{sec:eval:micro}) and macro-benchmarks (\S\ref{sec:eval:macro}).

\subsection{Experimental Setup}\label{sec:eval:hardware-setup}

We used a machine equipped with an Intel\textregistered{} Xeon\textregistered{} E-2288G CPU @ \(3.70\)\thinspace{}GHz ($8$ physical cores and activated hyper-threading) and \(128\)\thinspace{}GiB of RAM.
The system uses a \(2\)\thinspace{}TB Intel\textregistered{} 660p M.2 using PCIe 3.0\(x\)4 as persistent storage.
The operating system is Ubuntu 18.04.5 LTS with Linux 4.15.0-142 and \gls{sgx} software support is made available through the \gls{sgx} OOT driver in the most recent version 2.11 built from source~\cite{driver}.
We compare \sys{}'s performance with the master version of SCONE v5.3.0 as baseline.
\sys{}'s implementation is built on top of the same version of SCONE.

\subsection{Micro-benchmarks}\label{sec:eval:micro}
All micro-benchmarks have been executed with the Criterion benchmarking framework~\cite{criterion}.
The framework was configured to warm each benchmark for \(3\) seconds and afterward perform measurements for at least \(120\) seconds taking \(20\) samples.
Variance is omitted as it was negligible throughout all benchmarks.
We report the mean runtime as reported by Criterion.


\myparagraph{SHA-256 Computation Performance}\label{sec:eval:micro:sha256}
We investigate the overhead introduced by \sys{}'s interruptible SHA-256 implementation through evaluation of the hashing performance.
The benchmark hashes a buffer of various sizes and measures the throughput.
The buffer sizes correspond to differently sized \gls{sgx} enclaves, \ie{}, the size of an enclave program that must be hashed (measured) for attestation.

\begin{figure}[t]
    \centering
    \includesvg[height=.25\textwidth]{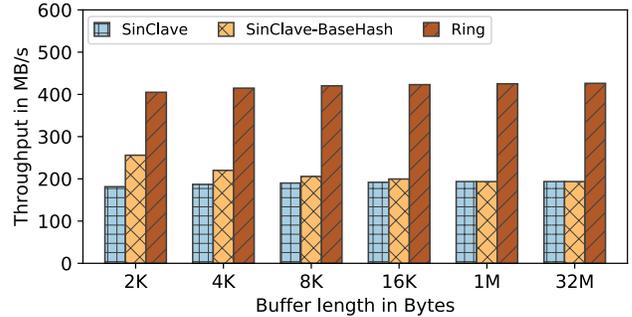}
    \caption{Calculation of a SHA256 checksum with different implementations.}\label{fig:eval:sha_comparison}
    \Description{A plot showing the evaluation of different implementations calculating a SHA256 checksum.}
\end{figure}

\begin{figure*}[!ht]
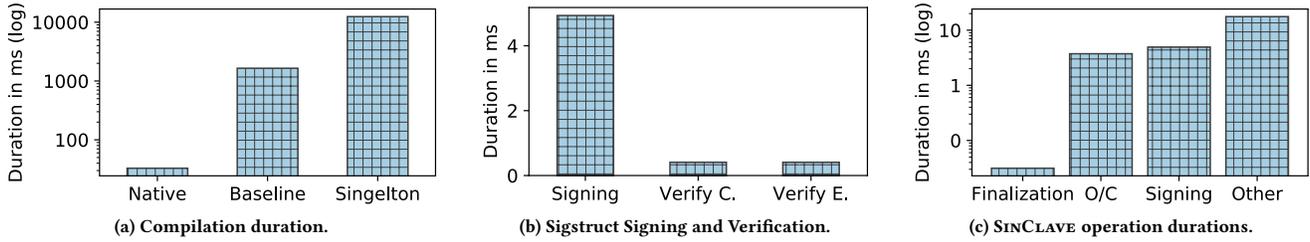

	\captionsetup[subfloat]{farskip=2pt,captionskip=1pt}
	
    \subfloat[Compilation duration.]
	{
        \includesvg[height=.167\textwidth]{figures/evaluation/compilation.svg}\label{fig:eval:compilation}
	}
	\hspace{3pt}
	\subfloat[Sigstruct Signing and Verification.]
	{
        \includesvg[height=.163\textwidth]{figures/evaluation/sigstruct.svg}\label{fig:eval:sigstruct}
	}
	\hspace{3pt}	
	\subfloat[\sys{} operation durations.]
	{
        \includesvg[height=.167\textwidth]{figures/evaluation/t2.svg}\label{fig:eval:cas_interface2}
	}
    \caption{Micro-benchmarks of \sys{}.}\label{fig:micro_bench}
    \Description{No description}
    \rk{Improve captions!}
\end{figure*}




Figure~\ref{fig:eval:sha_comparison} shows a comparison of three SHA256 checksum calculations.
The baseline for this comparison is Ring v0.16.20~\cite{ring}, a crypto library for Rust that uses highly optimized code comparable to OpenSSL\footnote{A single run of \texttt{openssl speed sha256} reveals a speed of about \(480\)\thinspace{}MB/s for a \(16\)\thinspace{}KB buffer.}.

\sys{}-BaseHash indicates the implementation of SHA-256 that produces a base enclave hash, \ie{}, the SHA-256 computation is interrupted before finalization of an enclave, and the Base Enclave Hash is encoded instead of finalizing the SHA computation.
\sys{} indicates the variant of \sys{}'s SHA-256 implementation without interruptions, \ie{}, no base enclave hash is calculated but the SHA-256 computation is finalized directly.

The results show that SHA-256 implementation in the Ring library outperforms the interruptible variant of \sys{} for all buffer sizes.
Independent of the buffer size it yields about \(405\)\thinspace{}MB/s.
The \sys{} implementation, which is directly comparable to Ring, yields about \(180\)\thinspace{}MB/s.
This overhead is the result of \sys{} using less optimized code.
\sys{}-BaseHash basically shows an equivalent performance compared to \sys{} for sufficiently large buffer sizes.
For small buffers, the overhead introduced by the additional encoding is significantly smaller than the effort necessary for SHA-256 finalization and, thus, \sys{}-BaseHash performs better with about 245~MB/s for 2~KB buffers.

Lastly, we measured the time it takes to finalize an enclave base hash into an enclave measurement.
This time is not influenced by the buffer size and requires constant \(32\)\thinspace{}\(\mu{}\)s.

\myparagraph{Enclave Program Compilation}\label{sec:eval:micro:compilation}
In SCONE, during compilation, the enclave's \gls{sigstruct} is embedded into the binary's ELF.
The enclave measurement, which is part of the \gls{sigstruct}, is calculated by the SCONE signing tool using SHA-256.
For \sys{} we adapted the algorithm of SHA-256 to support interruption and to be able to measure individual enclave code based on a common base enclave (\S{}~\ref{sec:protection}).
We evaluate the overhead introduced by our changes by measuring how long it takes to compile a small C program that only contains a return statement in the main function.
We compare our results to compilation with the system C compiler (native), and SCONE (baseline).

As shown in Figure~\ref{fig:eval:compilation}, baseline and \sys{} introduce significant overhead.
While native compilation takes only about \(0.033\)\thinspace{}s, baseline requires \(1.52\)\thinspace{}s and \sys{} \(6.26\)\thinspace{}s.
\sys{} needs about \(4\) times longer for compilation, while the ratio in the SHA-256 benchmark is just about \(2.25\).
Apart from the SHA-256 algorithm, there are no significant changes \sys{} introduces to the signer tool but
compared to the SHA-256 benchmark for which a continuous buffer is hashed at once, \ie{}, with a single call, the signer tool updates the SHA-256 computation iteratively.
Thus, we attribute this increased overhead to the less optimized code handling the entering and exiting of the hash computation.

\myparagraph{SigStruct Signing and Verification}\label{sec:eval:micro:sig_struct_signing}
In \sys{} each enclave's measurement is unique and only known at execution time.
\gls{sgx}, on the other hand, requires a \gls{sigstruct} that is matching the enclave measurement.
Thus, in \sys{} individual \gls{sigstruct} must be created for each singleton enclave that is started, which is done by the verifier, \ie{}, SCONE \gls{cas} (\S{}\ref{sec:sigstruct}).
\gls{sgx} \glspl{sigstruct} are signed with an RSA key with \(3072\)\thinspace{}bits.
We assess the introduced overhead by measuring the time it takes to sign the \gls{sigstruct}.
During On-Demand \gls{sigstruct} generation, the verifier must assess the authenticity of received \glspl{sigstruct} to prevent tampering, therefore we also measure the time it takes to verify a \gls{sigstruct}.

As shown in Figure~\ref{fig:eval:sigstruct}, \gls{sigstruct} signing takes \(4.9\)\thinspace{}ms on average.
While verification takes \(0.4\)\thinspace{}ms on average (``Verify C.'' in Figure~\ref{fig:eval:sigstruct}).
We further measure the time the verification takes in case it fails ``Verify E.''. 
The results show no difference to correct verification.

\myparagraph{Singleton Page Retrieval}
In this experiment, we measure the entire overhead introduced by singleton page retrieval.
Singleton page retrieval is the most distinct modification to the attestation procedure since this interaction is introduced by \sys{}.
Thus, we expect it to dominate the overhead introduced by \sys{}.

We measured the entire time it takes to retrieve a singleton page from the verifier, as well as, the time it takes to connect to the verifier and terminate the connection again after a simple no-op request.
Figure~\ref{fig:eval:cas_interface2} shows the results of these measurements split into individual operations necessary to complete the retrieval.
We use measurements from section \S{}\ref{sec:eval:micro:sha256} to attribute the latency.


The entire operation adds about \(26.3\)\thinspace{}ms to each enclave's start.
During this operation, a negligible amount of time is spent to verify the received \gls{sigstruct}'s integrity (\(0.4\)\thinspace{}ms) and to calculate the expected singleton's enclave measurement (\(32\)\thinspace{}\(\mu{}\)s).
Establishing the connection (O/C) requires \(3.74\)\thinspace{}ms and the signing of the singleton's \gls{sigstruct} \(4.93\)\thinspace{}ms.
The majority of time is spent with miscellaneous other necessary activities in the SCONE \gls{cas}.
These involve the loading and parsing of the configuration details from the encrypted database, as well as, the enforcement of access policies.

\begin{figure*}[!t]
	\centering
	\begin{minipage}{.48\textwidth}
	\vspace{-3pt} 
    \includesvg[width=.97\textwidth]{figures/evaluation/execution2.svg}
    \caption{Measurement of program execution.}\label{fig:eval:execution}
    \Description{No description.}
	\end{minipage}
    \hspace{15pt}
	\begin{minipage}{.48\textwidth}
    \vspace{3pt} 
    \includesvg[width=.98\textwidth]{figures/evaluation/macro-bench3_robert.svg}
    \caption{The performance overhead of \sys{} with real-world workloads.}\label{fig:macro-benchmarks}
    \Description{No description}
	\end{minipage}%
\end{figure*}

	
	

\myparagraph{Enclave Execution}
In our final micro-benchmark, we evaluate the overhead \sys{} introduces for running attested \gls{sgx} enclaves in total.
For this benchmark, we use the minimal C program from Section~\ref{sec:eval:micro:compilation} and measure its execution time under various configurations.
We ran it in simulation mode, \ie{}, without hardware protection, with \gls{sgx} hardware protection, and with hardware and attestation.
Also, we vary the enclave heap size from \(32\)\thinspace{}MB to \(2\)\thinspace{}GB.
We compare the individual measurements with each other to estimate the overheads introduced by the framework, the hardware protection, and the attestation procedure.

Figure~\ref{fig:eval:execution} shows the results for \sys{} and baseline.
We find that, throughout the benchmarks, \sys{} is slower only within single digit microsecond values compared to the baseline for execution modes that make no use of attestation.
Thus, we conclude that \sys{} does not introduce significant overhead to the framework's nor the hardware protection's overheads.

Attestation overhead increases slightly with enclave heap size, which is an effect of increased cache pressure.
For attested execution \sys{} introduces \(132\)\thinspace{}ms to \(144.2\)\thinspace{}ms compared to \(36.3\)\thinspace{}ms to \(65.9\)\thinspace{}ms introduced by the baseline system.
Compared to the overall execution time of up to \(5.2\)\thinspace{}s (min: \(258.1\)\thinspace{}ms), these overheads are only significant in experiments with small enclave heap size.

\subsection{Macro-benchmarks}\label{sec:eval:macro}

In this section, we evaluate our protection mechanism overhead using real-world applications including \emph{(1)} a simple python application with an encrypted volume (Python)~\cite{python-volume}, \emph{(2)} image classification using OpenVino~\cite{openvino} (OpenVino), and \emph{(3)} Pytorch CIFAR-10 dataset training~\cite{pytorch-cifar10} (Pytorch).
We measure the duration of these workloads when running them inside \gls{sgx} enclaves with \sys{} using the SCONE attestation service \gls{cas}.
We then enable the proposed protection mechanism \sys{} in SCONE and perform the same experiments as baseline to evaluate the impact of the protection mechanism on performance as shown in Figure~\ref{fig:macro-benchmarks}. 
\sys{} introduces \(1.03\)\%, \(2.49\)\%, and \(13.2\)\% overheads in the benchmarks for Python, Openvino, Pytorch, in comparison to the baseline.
Our benchmarks show that for real-world applications the overhead measured during the micro-benchmarks is negligible.

\section{Related work}

\myparagraph{Attacks against remote attestation}
The remote attestation in several TEEs is currently vulnerable to \emph{relay attacks}.
In these attacks, an adversary can control the OS (or other software) on the target host/device to relay incoming attestation requests to another host/device.
For example, in the context of TPM attestation, the relay attack is called \emph{cuckoo attack}~\cite{parno_bootstrapping}. 
In this attack, the adversary uses malicious software on the target host to redirect the attestation to a host that he controls.
Thus, the provisioning of confidential data is directed to the host controlled by the adversary.
To handle this attack, several defense mechanisms have been proposed however they still contain several limitations~\cite{fink_catching_2011}.
For example, they require dedicated measurement conditions, e.g., turn off of some OS processes. In addition, it has a high amount of false positives.

Meanwhile, in another approach Flicker~\cite{flicker2008} attests to only a single application at the time, requiring to shut down the entire OS and all other services and also requires a trusted administrator.
Recently, several attacks against Intel \gls{sgx} remote attestation have been presented:
ProximiTEE~\cite{proximi-tee} presents a relay attack on SGX attestation, or SGAxe~\cite{van2020sgaxe} shows an attack that bypasses the SGX remote attestations.
However these attacks require to first perform micro-architectural attacks such as side-channel attacks~\cite{Kocher2018spectre, chen2018sgxpectre, Foreshadow, Lipp2018meltdown}, thereafter use the gained secrets to attack the remote attestation.
In this paper, we present an attack against TEEs remote attestation without requiring to perform any microarchitectural attacks.

\myparagraph{Side-channel attacks}
Researchers recently demonstrated that TEEs are vulnerable to side-channel attacks~\cite{Kocher2018spectre, chen2018sgxpectre, Foreshadow, Lipp2018meltdown}.
To mitigate these attacks, protection mechanisms have been proposed~\cite{DRSGX, oblivious-ram2, oblivious-ram1, cloak, raccoon, szefer_cache_attacks}.
They follow the idea of hiding the memory access pattern of applications running inside SGX enclaves~\cite{cloak, oblivious-ram2, oblivious-ram1, DRSGX} by generating random memory accesses or shuffling the encrypted memory elements in a way that the observable access patterns are similar to each other and independent from the original access patterns.
Intel has been also working to release various patches, microcode updates to mitigate SGX side-channel attacks.
In this paper, we however turn our focus to the second key feature of \glspl{tee}, which is the \emph{remote attestation} mechanism.
We present an attack that bypasses the remote attestation protocol. Our attack does not require attacking the sealing mechanism of TEEs as the previous work~\cite{van2020sgaxe}. 

\myparagraph{Support legacy applications} 
Several SGX platforms~\cite{arnautov2016scone, priebe2019sgx, tsai2017graphene, occlum} have been developed to support legacy applications running with Intel \gls{sgx} without changing their source code and are widely used in practice~\cite{arnautov2016scone}.
SCONE~\cite{arnautov2016scone} was the first framework that supports switchless system calls to improve the performance of applications running with Intel SGX.
It supports porting many confidential cloud-native applications, including data analytics systems~\cite{perun,securetf,sgx-pyspark,secfl}, storage systems~\cite{avocado,pesos}, key management systems~\cite{palaemon}, and performance monitoring~\cite{teemon}. 
Following this idea, other SGX frameworks such as Graphene-SGX~\cite{tsai2017graphene}, SGX-LKL~\cite{priebe2019sgx}, or Occlum~\cite{occlum} have been proposed. 
Unfortunately, Graphene-SGX and Occlum only provide the building blocks for users to perform remote attestation and secret provisioning for their applications. 
SCONE and SGX-LKL support a transparently performing remote attestation mechanism for users. 
However, this paper demonstrates that the supported mechanism is not secure by showing an attack to bypass the remote attestation. 
We also provide novel protection against this attack and strengthen the remote attestation mechanism of SCONE with only lightweight overhead.

\section{Conclusion}
In this paper, we show that the measurement for remote attestation in TEEs is currently vulnerable to certain ``resue'' attacks of enclaves since it only guarantees software integrity before it runs, but not during runtime.
We demonstrate a such practical attack against different TEE frameworks. 
In addition, we provide a solution to strengthen the measurement by considering all aspects that determine the behavior of remote applications and ensure the measurement is fresh. 
We evaluate our protection mechanism with real-world applications. 
The evaluation shows that it can protect against reuse attacks and incurs only negligible overhead, which is only from $1.03$\% to $13.2$\% depending on running applications.

\myparagraph{Acknowledgements}
We thank Andrey Brito, our reviewers, and our shepherd Valerio Schiavoni for their work.
This publication was funded by the European Union's Horizon Europe
research and innovation program under grant agreements 101092646 (Cloudskin), 101092644 (Neardata), 101016577 (AI-Sprint), by the Deutsche Forschungsgemeinschaft (DFG) in subproject B07 of the Collaborative Research Center Transregio 96 (project 174223256), and by the Federal Ministry of Education and Research (6G-Life).

\bibliographystyle{ACM-Reference-Format}
\bibliography{biblio}

\end{document}